\documentclass[prx,twocolumn,floatfix,superscriptaddress,longbibliography]{revtex4-1}
\usepackage{tikz}
\usepackage{algpseudocode}
\usepackage{appendix}
\usepackage[ruled]{algorithm}
\usepackage{mathtools}
\usepackage{multirow}
\usetikzlibrary{shapes,arrows,positioning}
\usepackage{appendix}
\usepackage{standalone}
\usepackage{amssymb,amsmath,amstext}
\usepackage{graphicx}
\usepackage{epstopdf}
\usepackage{color}
\usepackage[T1]{fontenc}
\usepackage{bbold}
\usepackage{bbm}
\usepackage{float}
\usepackage{latexsym}
\usepackage{xr-hyper}
\usepackage[colorlinks=true,citecolor=blue,linkcolor=magenta]{hyperref}
\usepackage[latin1]{inputenc}
\usepackage{hhline}
\usepackage{physics}

\usepackage[normalem]{ulem}
\usepackage[final]{pdfpages}
\usepackage{forloop}

\usetikzlibrary{shapes,arrows}
\long\def\ca#1\cb{} 


   \makeatletter
     \renewcommand\@make@capt@title[2]{%
      \@ifx@empty\float@link{\@firstofone}{\expandafter\href\expandafter{\float@link}}%
       {\textbf{#1}}\@caption@fignum@sep#2\quad}%
     \makeatother
\makeatletter 
\renewcommand{\fnum@figure}{\textbf{Figure~\thefigure}}
\makeatother

 \usepackage[compact]{titlesec}
    \titlespacing{\section}{0pt}{4ex}{2ex}
    \titlespacing{\subsection}{0pt}{4ex}{3ex}
    \titlespacing{\subsubsection}{0pt}{3ex}{2ex}

\titleformat{\section}
  {\normalfont\fontsize{10}{12}\bfseries\filcenter}{\thesection.}{1em}{}

\titleformat{\subsection}
  {\normalfont\fontsize{9}{11}\bfseries\filcenter}{\thesubsection.}{1em}{}

\makeatletter
\AtBeginDocument{\let\LS@rot\@undefined}
\makeatother

\begin{document}

\title{Correlation-Informed Permutation of Qubits for Reducing Ansatz Depth in VQE}

\author{Nikolay V. Tkachenko}
\thanks{These authors contributed equally.}
\affiliation{Chemistry Division, Los Alamos National Laboratory, Los Alamos, New Mexico 87545, USA}

\author{James Sud}
\thanks{These authors contributed equally.}
\affiliation{Department of Physics, University of California, Berkeley, CA 94720, USA}

\author{Yu Zhang}
\email{zhy@lanl.gov}
\affiliation{Theoretical Division, Los Alamos National Laboratory, Los Alamos, New Mexico 87545, USA}

\author{Sergei Tretiak}
\affiliation{Theoretical Division, Los Alamos National Laboratory, Los Alamos, New Mexico 87545, USA}

\author{Petr M. Anisimov}
\affiliation{Accelerators and Electrodynamics Group, Los Alamos National Laboratory, Los Alamos, New Mexico 87545, USA}

\author{Andrew T. Arrasmith}
\affiliation{Theoretical Division, Los Alamos National Laboratory, Los Alamos, New Mexico 87545, USA}

\author{Patrick J. Coles}
\affiliation{Theoretical Division, Los Alamos National Laboratory, Los Alamos, New Mexico 87545, USA}

\author{Lukasz Cincio}
\email{lcincio@lanl.gov}
\affiliation{Theoretical Division, Los Alamos National Laboratory, Los Alamos, New Mexico 87545, USA}

\author{Pavel A. Dub}
\email{pdub@lanl.gov}
\affiliation{Chemistry Division, Los Alamos National Laboratory, Los Alamos, New Mexico 87545, USA}

\begin{abstract}
The Variational Quantum Eigensolver (VQE) is a method of choice to solve the electronic structure problem for molecules on near-term gate-based quantum computers. However, the circuit depth is expected to grow significantly with problem size. Increased depth can both degrade the accuracy of the results and reduce trainability. In this work, we propose a novel approach to reduce ansatz circuit depth. Our approach, called PermVQE, adds an additional optimization loop to VQE that permutes qubits in order to solve for the qubit Hamiltonian that minimizes long-range correlations in the ground state. The choice of permutations is based on mutual information, which is a measure of interaction between electrons in spin-orbitals. Encoding strongly interacting spin-orbitals into proximal qubits on a quantum chip naturally reduces the circuit depth needed to prepare the ground state. For representative molecular systems, LiH, H$_2$, (H$_2$)$_2$, H$_4$, and H$_3^+$, we demonstrate for linear qubit connectivity that placing entangled qubits in close proximity leads to shallower depth circuits required to reach a given eigenvalue-eigenvector accuracy. This approach can be extended to any qubit connectivity and can significantly reduce the depth required to reach a desired accuracy in VQE. Moreover, our approach can be applied to other variational quantum algorithms beyond~VQE.
\end{abstract}

\maketitle

\section{Introduction}

Quantum computing is expected to revolutionize computational chemistry by achieving polynomial scaling in both the number of quantum particles and the quality of the description of the system (e.g., number of orbital basis functions or numerical grid points)~\cite{cao2018quantum,mcardle2020quantum}. The Variational Quantum Eigensolver (VQE) has emerged as a viable algorithm~\cite{Peruzzo2014} to find asymptotically exact lowest eigenvalues for solutions to the Schr{\"o}dinger equation on Noisy Intermediate-Scale Quantum (NISQ) devices. VQE-based ground state electronic energy calculations of small molecular systems (e.g., H$_2$, BeH$_2$, H$_2$O, alkali metal hydrides and H$_{12}$) in minimal basis sets (e.g., contracted Gaussians sto-ng family) were experimentally implemented using superconducting circuits~\cite{o2016scalable,Kandala2017,kandala2019error,RN125,arute2020science} and trapped ions~\cite{nam2020ground} as physical qubits. As the age of quantum supremacy dawns~\cite{arute2019quantum}, demonstrating quantum advantage for chemistry will naturally involve considering larger molecules and/or basis sets. In addition to quantum circuit width (number of qubits), this will, in turn, increase the quantum circuit depth, which typically grows polynomially in the problem size~\cite{Lee2019}.

The growing circuit depth with problem size causes two main issues. One issue is the accumulation of hardware noise, which impacts the accuracy of the results. The other issue is the trainability of the ansatz parameters, i.e., whether the parameters have large enough gradients to allow for progress in the optimization. This concern arises since increased circuit depth leads to smaller gradients~\cite{mcclean2018barren,cerezo2020cost,sharma2020trainability,cerezo2020impact}, and accurate estimation of small gradients on a quantum computer requires a large number of runs (or ``shots''). In fact, these two issues are related as the accumulation of hardware noise also leads to smaller gradients~\cite{wang2020noise}. This highlights the importance of keeping the circuit depth shallow in variational ans\"atze.

Some strategies partially address these problems, such as improved classical optimizers~\cite{kubler2019adaptive,arrasmith2020operator,Sung2020}, parameter initialization strategies~\cite{volkoff2020large,verdon2019learning,grant2019initialization}, error mitigation~\cite{li2017efficient,temme2017error,czarnik2020error}, and noise resilience~\cite{sharma2019noise}. On the other hand, the most direct strategy would be to somehow reduce ansatz circuit depth. Some promising approaches have been proposed for this purpose, largely focusing on adaptive ans\"atze~\cite{grimsley2018adapt,tang2019qubit,ryabinkin2018qubit,ryabinkin2020iterative}. Given that reducing ansatz depth will improve both the training complexity and the accuracy of VQE, it is a crucial research direction that could bring us closer to realize quantum advantage for electronic structure calculations. 

It is worth emphasizing that lack of complete qubit connectivity on NISQ devices \cite{10.1145/3297858.3304007} makes it necessary to increase the depth of quantum circuit. Thus, any efforts to minimize ansatz depth in VQE should account for the connectivity of the specific NISQ hardware. For instance, if two qubits are highly entangled in the exact solution of an electronic structure problem, yet the qubits are physically distant on a device with limited connectivity, a deeper circuit is required for accurate simulation. Long sequences of noisy two-qubit gates are necessary to entangle distant qubits, which can degrade the fidelity and reduce trainability. 

\begin{figure}[t]
    \centering
    \includegraphics[width=\columnwidth]{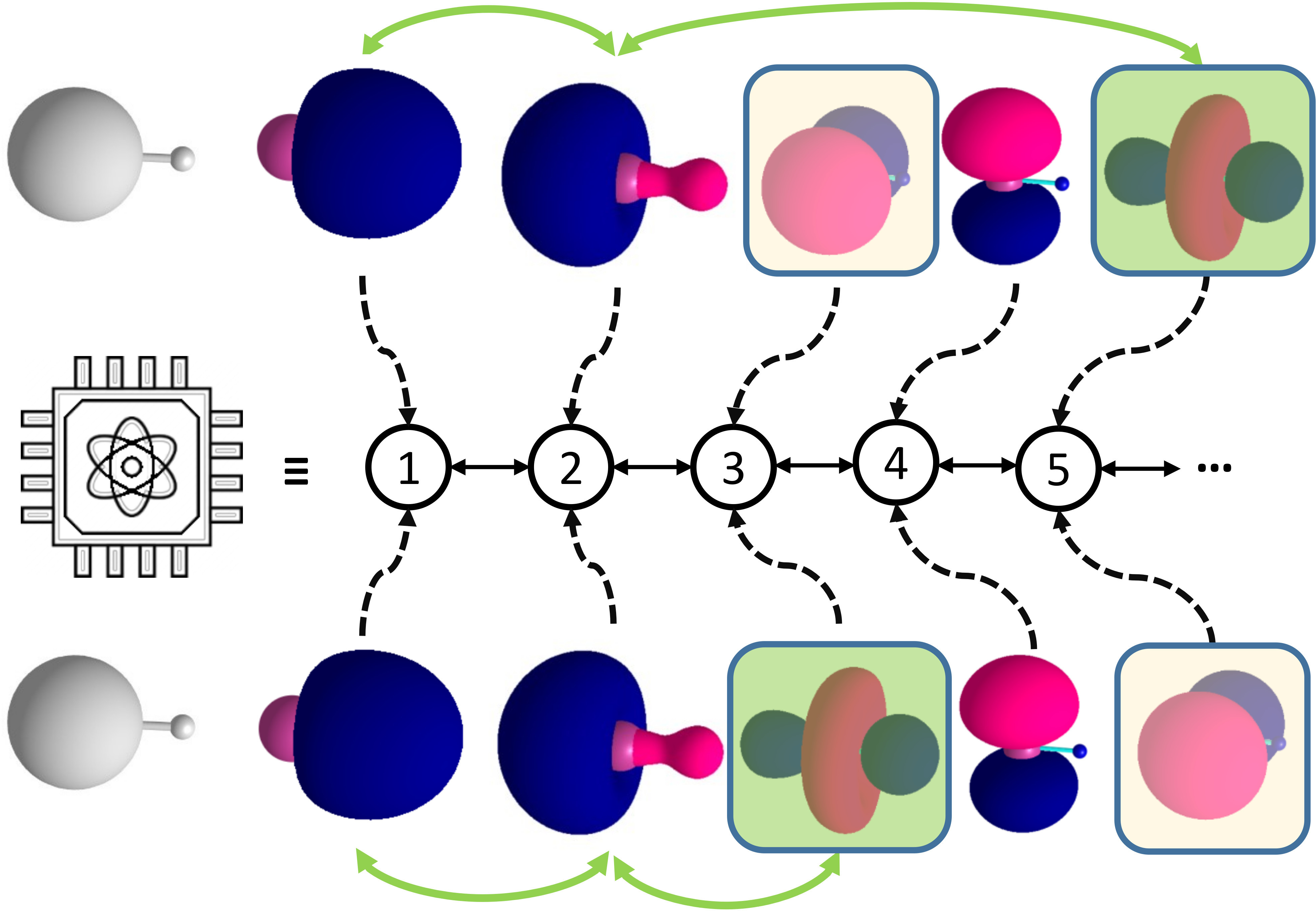}
    \caption{Correlation-informed permutation schematics on the example of LiH molecule in a reduced active-space of sto-3g basis: direct encoding of five $\alpha$ spin-orbitals into five physical qubits on a linear architecture quantum chip before (top) and after (bottom) permutation based on orbital entanglement information (green double-arrows show correlated orbitals). }
    \label{fig:Intro}
\end{figure}

In this work, we introduce a novel approach to reduce ansatz depth by permuting the pattern with which the Hamiltonian is embedded onto qubits. Specifically, we employ a correlation-informed approach, where the permutation is chosen based on the mutual information defining entanglement between two individual spin-orbitals \cite{Rissler2006,HUANG20051}. Our main idea is illustrated in Fig.~\ref{fig:Intro} on the example of selected spin-orbitals of LiH molecule. Entanglement of spin orbitals triggers permutation of the embedding to ensure that these orbitals are encoded into qubits that are placed physically close to each other on an actual quantum chip.

We numerically implement such qubits permutations assuming a linear qubit connectivity for the LiH, H$_2$, (H$_2$)$_2$, H$_4$, and H$_3^+$ molecular systems. In all cases, permutations significantly reduce the circuit depth required to reach a given energy accuracy, relative to the unpermuted case. To realize this permutation technique, we introduce PermVQE algorithm, an added layer on top of the original VQE posed in \citep{Peruzzo2014}. At a fixed circuit depth $L$, this algorithm, as depicted in Fig.~\ref{fig:algorithm} starts from the approximate ground state wave function of a given initial Hamiltonian, then uses it to calculate an optimal reordering of qubits and builds a permuted Hamiltonian. The algorithm iteratively repeats this process until an adequate reordering of qubits is obtained, and the best-permuted qubit Hamiltonian is used to variationally calculate the best-possible (within $L$) ground state wave function and associated energy.

\section{The PermVQE Algorithm}

\subsection{Overview}

A diagram of the PermVQE algorithm is shown in Fig.~\ref{fig:algorithm}. As in the standard VQE for electronic structure, the first step is the initial mapping of the fermionic Fock-space states as well as creation and annihilation operators into Hilbert-space states and Pauli operators of qubits based on one of established second-quantized encoding methods, \textit{vide infra}. This is followed by a standard VQE loop to variationally learn an approximate correlated ground state wave function at a fixed circuit depth $L$. The next step is to perform local tomography on two-qubit subsets in order to generate mutual-information matrix, which we call the entanglement map. This step also minimizes a cost function that quantifies the amount of long-range correlations (for the specific hardware's connectivity), while varying the qubit index labels. As a result we arrive at an optimal permutation of the qubits and hence a new Hamiltonian, which is then fed back to the VQE subroutine for another iteration. We now provide more details on these various subroutines.

\subsection{Initial qubit mapping}

In PermVQE, the first step is an initial spin-orbital to qubit mapping, subject to a chosen transformation~\cite{cao2018quantum,mcardle2020quantum}. As discussed below in Sec.~\ref{sec:Numerics}, the performance of PermVQE can be affected by the initial qubit mapping. In particular, it may be desirable to select an initial qubit mapping that leads to a more sparse entanglement map, i.e., minimizing the number of qubits being entangled. This would then allow for qubit permutations to have a more significant effect. In Sec.~\ref{sec:Numerics}, we compare three popular second quantized basis set encoding methods: Jordan-Wigner (JW) \cite{RN126}, Bravyi-Kitaev (BK)~\cite{BRAVYI2002210}, and Parity~\cite{parity2012,bravyi2017tapering}. We find that the Jordan-Wigner transformation facilitates sparser entanglement map, and hence improves performance of PermVQE.

\subsection{Entanglement map}

After obtaining the initial qubit Hamiltonian and running VQE under fixed circuit depth $L$, the next step of PermVQE is to produce an entanglement map reflecting electronic correlations in the approximate ground state. For this purpose, we calculate the quantum mutual information for all pairs of qubits, which provides a measure of the total correlation including both quantum and classical correlations. Previous results obtained for the orbital ordering problem in Density Matrix Renormalization Group (DMRG) method in classical quantum chemistry calculations \citep{Rissler2006} showed that the quantum mutual information is a reliable parameter to quantify the correlation between two quantum particles. The quantum mutual information between qubits $i$ and $j$ is defined as follows:

\begin{equation}\label{mutual}
    I_{ij}=\frac{1}{2}(S_i+S_j-S_{ij})(1-\delta_{ij})\,,
\end{equation}
where $S_i$ and $S_{ij}$ are the single-qubit and two-qubit von Neumann entropies, respectively. The Kronecker $\delta$ sets all diagonal elements $I_{ii}$ to zero. The single-qubit von Neumann entropy $S_i$ is given by:
\begin{equation}\label{entropy}
    S_i = - \Tr(\rho_i\log\rho_i) = -  \sum_\alpha \lambda_i^{(\alpha)}\log \lambda_i^{(\alpha)}\,,
\end{equation}
and $S_{ij}$ is defined analogously. Here $\rho_i$ is the reduced one-body density matrix of qubit $i$ and $\{\lambda_i^{(\alpha)}\}_\alpha$ are its eigenvalues.

\begin{figure}[t]
    \centering
    \includegraphics[width=\columnwidth]{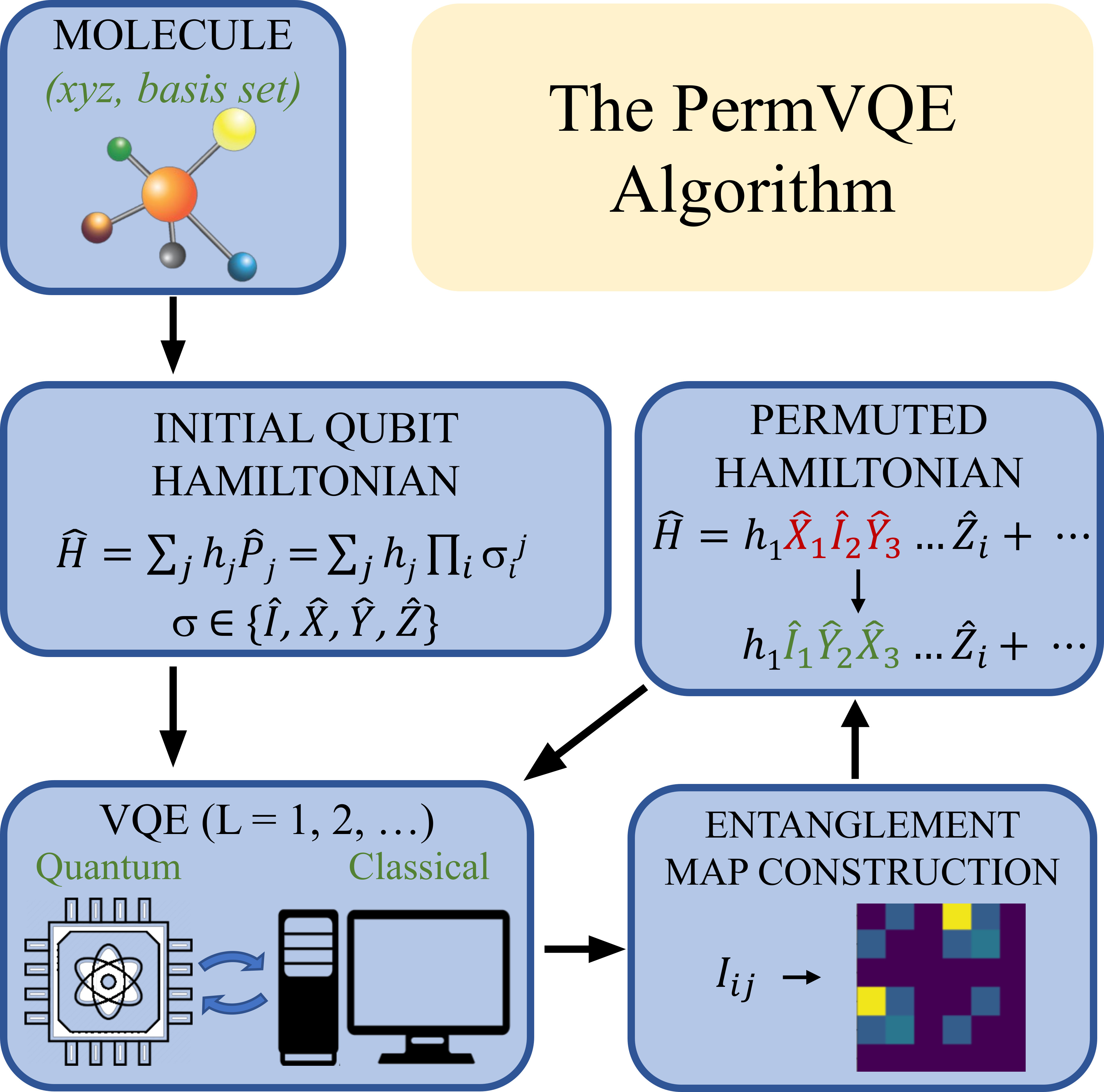}
    \caption{The PermVQE Algorithm. After obtaining approximate ground state wave function from the initial Hamiltonian at a fixed circuit depth $L$ through VQE, the entanglement map is produced based on mutual information $I_{ij}$. Minimization of a cost function (which quantifies the amount of long-range correlations) over possible permutations of qubits, returns the permuted Hamiltonian, which is then fed back into the VQE algorithm. This procedure is iteratively repeated until the best-permuted Hamiltonian is generated from which the best-possible (within $L$) ground state wave function and energy are obtained.}
    \label{fig:algorithm}
\end{figure}

We note that $\rho_{ij}$ (and its marginals $\rho_{i}$ and $\rho_{j}$) can be obtained from two-qubit tomography, which involves measuring 15 Pauli operators. There are $n(n-1)/2$ pairs of qubits, with $n$ being the number of qubits. Hence the number of operator measurements needed to construct the entanglement map is $15n(n-1)/2$. Some of these operators will commute and thus can be measured simultaneously.

Based on the mutual information values for each pair of qubits, we build an $n \times n$ matrix. This matrix $\mathbf{I} = \{I_{ij}\}$, called the entanglement map, is useful to illustrate the amount and the length-scale of the correlations in the approximate ground state. Moreover, we use it to define a cost function described next.

\subsection{Cost Function}

To quantify the amount of long-range correlations, we introduce a cost function as follows. For a NISQ device with a given connectivity, let $d_{ij}$ denote the distance between qubits $i$ and $j$, which can be precisely defined as the number of edges in the shortest path through the connectivity graph between these qubits. Alternatively, one can view $d_{ij}$ as the minimum number of swap gates (plus one) needed to make qubits $i$ and $j$ nearest neighbors. Then, for any qubit connectivity, a cost function can be defined as
\begin{equation}\label{eqcost}
    C(\mathbf{I}) = \sum_{i<j} f(d_{ij}) I_{ij}\,,
\end{equation}
where $f(\cdot)$ is a monotonously increasing function of $d_{ij}$. 
As an example, one could choose a power-law function: $f(d_{ij}) = d_{ij}^{\beta}$ for some $\beta > 0$. When $\beta =2$, we have 
\begin{equation}\label{eqcostbetaeq2}
    C(\mathbf{I}) = \sum_{i<j} d_{ij}^2 I_{ij}\,,
\end{equation}
being the choice of cost function employed in our numerics below. Specifically, we consider a linear connectivity in our numerics, in which case $d_{ij} = |i -j|$, and the cost function becomes $ C(\mathbf{I}) = \sum_{i<j} |i -j|^2 I_{ij}$. The choice of the cost function is dictated by the optimization procedure described below.

\subsection{Cost-Function Minimization}

We define a permutation $P$ as a bijection from the set of qubit indices to itself. The action of $P$ will affect the entanglement map $\mathbf{I}$, and we are interested in solving the optimization problem:
\begin{equation}\label{costopt}
P_{\text{opt}} = \arg \min_P C(P\mathbf{I}P^{-1})\,. 
\end{equation}

After solving this optimization problem, the PermVQE algorithm uses $P_{\text{opt}}$ to permute the qubit indices and produce a new Hamiltonian. For example, if the original Hamiltonian is $\widehat{H} = \widehat{X}_1 \widehat{Y}_2 \widehat{Z}_3$ and $P_{\text{opt}}$ transforms indices as follows: $1\mapsto3,  2\mapsto2,  3\mapsto1$, then we produce the permuted Hamiltonian $\widehat{H} = \widehat{X}_3 \widehat{Y}_2 \widehat{Z}_1$.

There are several methods suitable for minimization of $C(P\mathbf{I}P^{-1})$ 
over qubit permutations. In principle, one could take a brute-force approach. That is, one could explicitly construct the $n!$ different permutations and check which one of them produces the lowest value of $C(P\mathbf{I}P^{-1})$. 
While this approach will certainly find the best permutation for a given entanglement map, the computational cost grows factorially in $n$, and hence scalability to large problem sizes is problematic. Instead, we focus on a more practical technique, as follows.

\subsubsection{Spectral Graph Algorithm}

The mutual information matrix $\mathbf{I}$ can also be considered as a weighted graph where the weighted edge represents the mutual information between two qubits. In particular, the minimization of a quadratic cost function, $C(\mathbf{I})=\sum_{i<j}|i-j|^2I_{ij}$, can be related to spectral graph theory~\cite{Fiedler1975,Reiher2011pra}. For a given $\mathbf{I}$, we can define the graph Laplacian $\mathbf{L}$ as follows:
\begin{equation}
    \mathbf{L}=\mathbf{D}-\mathbf{I}, \quad D_{ij}=\delta_{ij} \sum_k I_{ik} \,.
\end{equation}

Spectral graph theory has shown~\cite{Fiedler1975} that the Fiedler vector, or the second smallest eigenvector of $\mathbf{L}$, is the solution that can provide low values of the following function:
\begin{equation}\label{fiedler}
    F(x)=x^{\dagger}\mathbf{L}x = \sum_{ij} I_{ij} (x_i-x_j)^2\,,
\end{equation}
which is exactly the cost function that we defined above. Sorting the entries of the Fiedler vector in either ascending or descending order provides the optimized ordering of qubits~\cite{Fiedler1975}. Note that the diagonalization of the Laplacian matrix $\mathbf{L}$ scales polynomially with number of qubits~$n$. Hence, the cost-function minimization based on the spectral graph algorithm provides an efficient way of finding optimal qubit ordering, with polynomial scaling in $n$.

\section{Numerical Implementations}\label{sec:Numerics}

\subsection{Ising Toy Models with \textit{Exact} Wave Functions}

To demonstrate the effect of permutations on circuit depth required to reach exact solution in the VQE, we start with toy models and exact wave functions giving rise to exact entanglement maps (here only one loop for qubits reordering is needed). For that purpose we analyze several 6-qubit model Ising Hamiltonians with artificially engineered entanglement. The generic Ising Hamiltonian is given by:

\begin{equation}\label{Ising Hamiltonian}
    \widehat{H} = \sum_i \hat{Z}_i + \sum_{ij} C_{ij}\hat{X_i}\hat{X_j} + \sum_{ijk} C_{ijk}\hat{X_i}\hat{X_j}\hat{X_k} + ...\,.
\end{equation}

\begin{table}[]
\begin{tabular}{|l|c|c|}
\hline
\multirow{2}{*}{\textbf{Hamiltonian}} & \multicolumn{2}{p{4cm}|}{\textbf{Depth required to reach exact solution}} \\ \cline{2-3} 
                  &    \textbf{Not\hspace{3pt}  \newline permuted}       &     \textbf{Permuted}      \\ \hline
$\widehat{H}_1=\sum\limits_{i}\widehat{Z}_i  + \widehat{X}_1\widehat{X}_6$              &    7       &    1       \\ \hline
$\widehat{H}_2=\sum\limits_{i}\widehat{Z}_i  + \widehat{X}_1\widehat{X}_6 + \widehat{X}_2\widehat{X}_5$    &      8     &    1       \\ \hline
$\widehat{H}_3=\sum\limits_{i}\widehat{Z}_i  + \widehat{X}_1\widehat{X}_3\widehat{X}_5$  &      4     &    2       \\ \hline
$\widehat{H}_4=\sum\limits_{i}\widehat{Z}_i  + \widehat{X}_1\widehat{X}_6 + \widehat{X}_1\widehat{X}_5$ &    8       &     3      \\ \hline
$\widehat{H}_5=\sum\limits_{i}\widehat{Z}_i  +2* \widehat{X}_1\widehat{X}_2\widehat{X}_5\widehat{X}_6$ &  5         &   5        \\ \hline

\end{tabular}
\caption{\label{tab:table-toy-model}Ansatz depth for 6-qubit Ising models. For various Ising models, the ansatz depth that is required to reach the exact energy is shown for the not-permuted case and for the optimally permuted case. 
}
\end{table}

\begin{figure}[t]
    \centering
    \includegraphics[width=8cm]{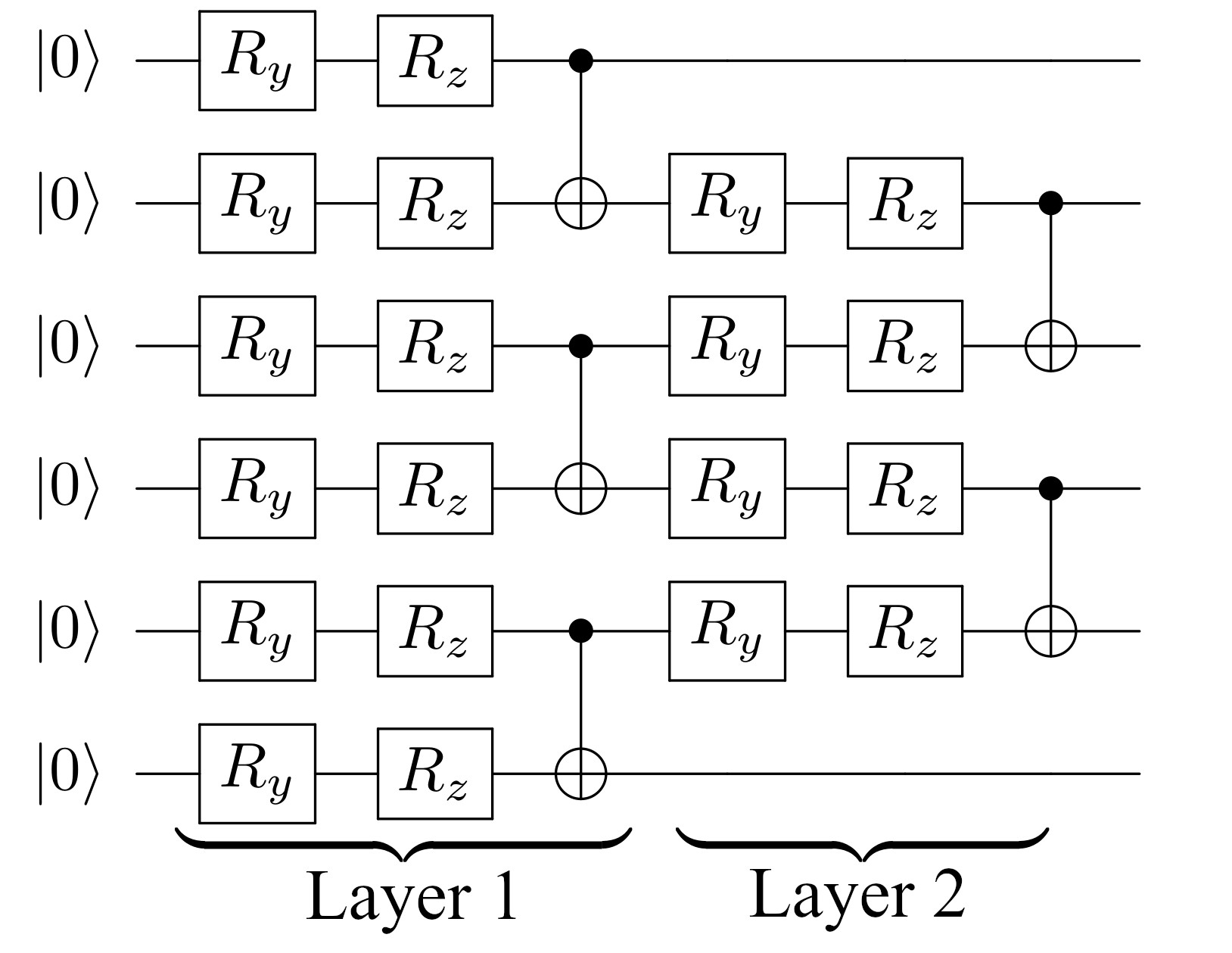}
    \caption{Illustration of the 2-Depth RyRz Ansatz that was used for VQE calculations using model Ising Hamiltonians.}
    \label{fig:ising_depth_2}
\end{figure}
 
We choose the heuristic two-layer RyRz ansatz shown in  Fig.~\ref{fig:ising_depth_2}, as this hardware-efficient ansatz is naturally adapted for NISQ devices to ensure the nearest neighbor connectivity between qubits.
The ansatz circuit with $L$ layers is denoted as an ``$L$-Depth'' circuit. We consider five different model Hamiltonians, given in Table~\ref{tab:table-toy-model}. The entanglement maps for those Hamiltonians are shown in Fig. S1 in Supplemental Material (SM).
From a comparison of different entanglement maps, we notice that the entanglement is more localized in the permuted case supporting the choice of the cost-function defined above. We find that for the majority of considered Hamiltonians, qubit permutations significantly reduce the number of ansatz layers required to reach the exact energy of the system as shown in Table~\ref{tab:table-toy-model}. For the case of the simplest $\widehat{H}_1$ and $\widehat{H}_2$ Hamiltonians, the exact solution is already found at $L = 1$, where only the distinct qubit pairs are entangled.  As expected, larger depth further reduces the advantage of permutations since long circuit depth allows to solve the problem exactly for the unpermuted (default) Hamiltonian. Thus, we expect that qubit permutations could be an ideal method to variationally learn the ground state wave functions on NISQ devices where circuit depths are limited.

For each Hamiltonian, we also compare the difference in permutated vs unpermuted  cost-function values to the difference in permuted vs unpermuted depth required to converge to the true ground state, see Table S1 in SM. Notably, a larger difference in cost-function appears to be concomitant with a larger difference in the required ansatz depth, which may indicate a correlation between these parameters. However, for a large number of entangled qubits (e.g. $\widehat{H}_5$), this correlation fails. Nevertheless, the cost-function difference can be used as an initial estimate of the possible permutations.

\begin{figure}[t]
    \centering
    \includegraphics[width=\columnwidth]{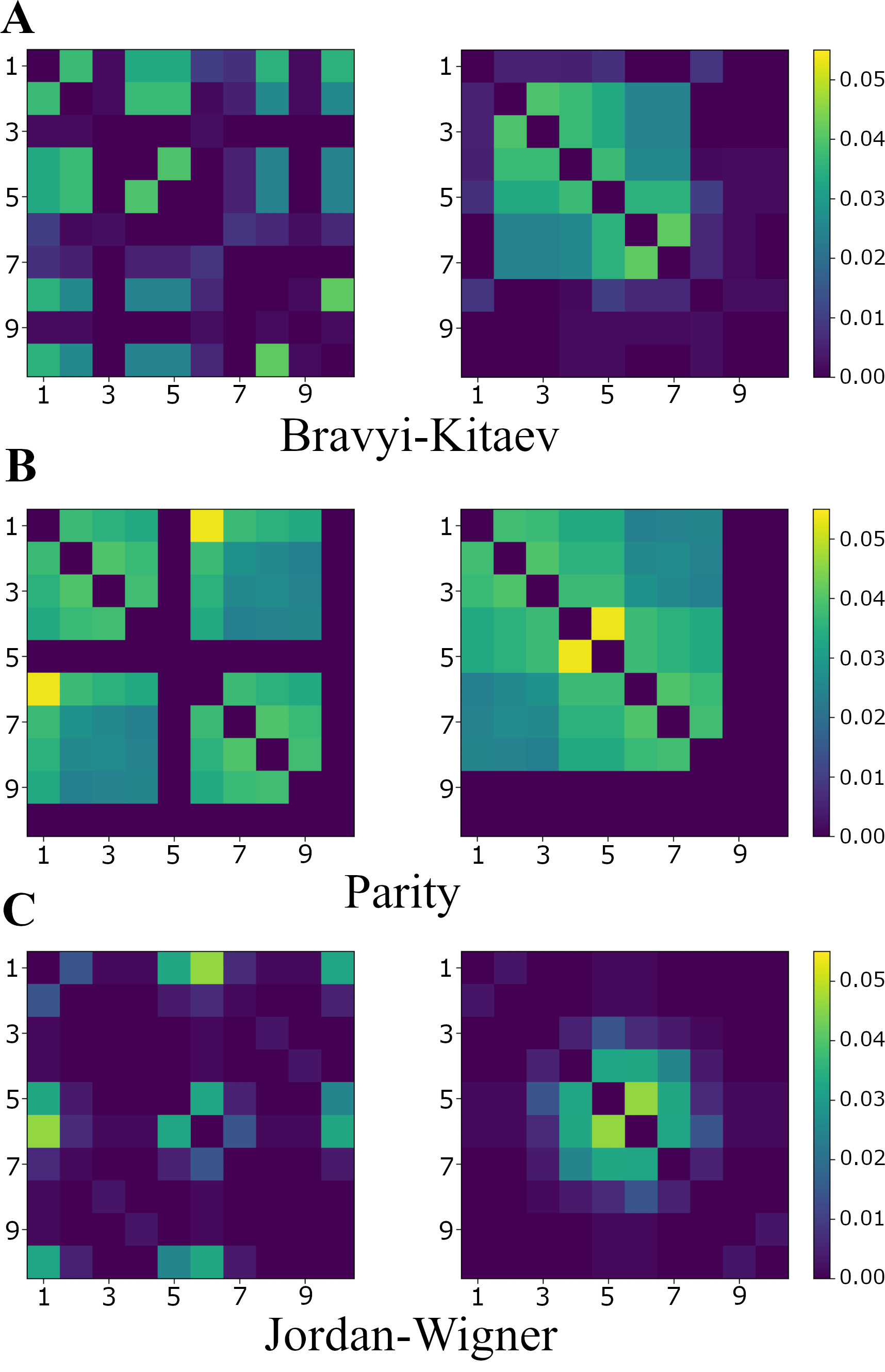}
    \caption{Entanglement maps built from the \textit{exact} wave function for LiH/sto-3g in a reduced active space of 10 spin-orbitals for the three commonly used mappings (labeled A, B and C). Left column: default Hamiltonian used, Right column: permuted Hamiltonian used. The numbers on the axes indicate qubit indices. Different values of mutual information $I_{ij}$ are shown with different colors with the color map shown on the right.} 
    \label{fig:lih_mappings}
\end{figure}

\subsection{Molecular Systems with \textit{Exact} Wave Functions}
For the actual molecular systems, the entanglement maps are expected to be vastly more complicated. Moreover, the type of mapping of fermionic-to-qubit Hamiltonian can also affect the entanglement map of the system. To demonstrate proof-of-principle advantage of qubit permutations on a circuit depth required to reach near-to-exact solution in the VQE of chemical systems, we consider five molecular LiH, H${}_2$, $(\mathrm{H}_2)_2$ T-shaped Van der Waals complex, H${}_4$ third-order saddle point and $\mathrm{H}_3^+$ cyclic cation systems, which were investigated in ideal environment (exact wave function, exact entanglement map, no noise consideration). Here, we compare $\Delta E = E_\mathrm{VQE} - E_\mathrm{exact}$ energy error as a function of ansatz depth $L$, where $E_\mathrm{VQE}$ is converged variational energy when using unpermuted or permuted Hamiltonian in the best identified (JW) encoding (see below), and $E_\mathrm{exact}$ is exact energy in the full or reduced active space obtained by direct Hamiltonian matrix diagonalization in the basis of Slater Determinants of spin-orbitals. For all systems a full active space of sto-3g basis set \citep{Pople1969} was used, except LiH and H${}_2$ molecules, for which a reduced active space of 10 spin-orbitals and a full active space of 6-31G basis set \citep{Pople1971} were selected, respectively. Slightly-modified versions of hardware-efficient RyRz and Ry ans\"atze \cite{RN1288} were used for all the molecules (see Figs. S2-S3 in SM), except $\mathrm{H}_3^+$ for which a particle-preserving ansatz was employed, \textit{vide infra}. For all these systems, we present graphs of converged VQE energy with and without permutations in the main text.

\subsubsection{Initial Qubit Mapping}

\begin{figure*}[t]
    \centering
    \includegraphics[width=\textwidth]{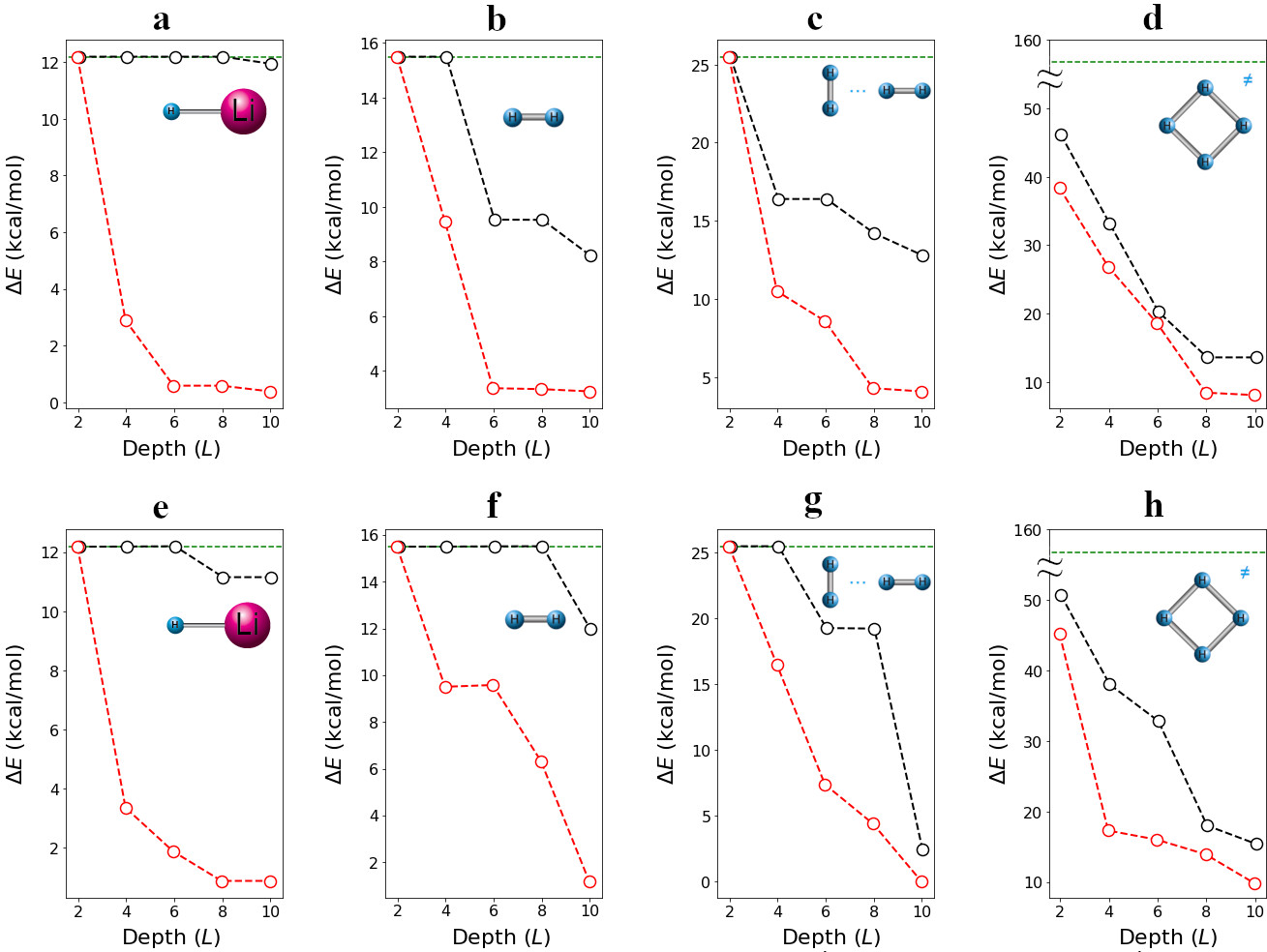}
    \caption{Proof-of-principle energy error $\Delta E = E_\mathrm{VQE} - E_\mathrm{exact}$ versus ansatz depth $L$ for various molecular species based on default (black) and permuted qubit Hamiltonians (red) built from the \textit{exact} wave functions. Top (a-d) and bottom (e-h) panels show the cases for the hardware-efficient RyRz and Ry ans\"atze, respectively. Green dotted-lines correspond to \textit{pseudo}-correlation energy $\Delta E_\mathrm{pc} = E_\mathrm{HF} - E_\mathrm{exact}$.}
    \label{fig:all_plots}
\end{figure*}

There are multiple choices of spin-orbital to qubit mappings~\cite{cao2018quantum,mcardle2020quantum}. For electronic structure problem, three popular second quantized encoding basis set methods Jordan-Wigner (JW) \cite{RN126}, Bravyi-Kitaev (BK)~\cite{BRAVYI2002210}, and Parity~\cite{parity2012,bravyi2017tapering} primarily differ in the pattern to store information on the occupation number, the parity and the number of qubit operations for simulating fermionic creation or annihilation operators. While it was shown that the BK mapping can reduce the number of operations to $O(\log_2(N))$ for simulating each fermionic operator \cite{RN127}, entanglement maps in this encoding contain a larger number of entangled qubits, which may reduce the advantage of permutations (Fig.~\ref{fig:lih_mappings}A). The same behavior was observed for the Parity mapping (Fig.~\ref{fig:lih_mappings}B). The reason for this
is likely relevant to the fact that both BK and Parity store information on the occupation number nonlocally, thus delocalizing the entanglement. In contrast, the occupation number-locality preserving JW mapping produces sparse entanglement map (Fig.~\ref{fig:lih_mappings}C). We expect that the advantage of permutations will be more pronounced in this mapping, which is further used as default throughout this paper. For example, LiH in the JW mapping includes 6 entangled qubits unlike the BK and Parity mappings in which 8 qubits are entangled, see Fig.~\ref{fig:lih_mappings}. The comparison of JW entanglement maps for other molecular systems is shown in Fig. S4 in SM. We also compare exact entanglement maps for qubit Hamiltonians generated with IBM's Qiskit \citep{Abraham2019} and Google's Openfermion \citep{McClean2017} frameworks for the LiH molecule in Fig.~S5 in SM. 
Entanglement maps are identical in JW encoding after qubit permutations, while they are different for BK and Parity mappings. This result is expected taking into account that Qiskit orders (by default) spin-orbitals based on spin (with ``up'' first and ``down'' next), whereas Openfermion performs even-odd ordering.

\subsubsection{$\mathrm{LiH}$ Molecule}
For the analysis of LiH in sto-3g basis, the lowest-energy spatial orbital was assumed to be doubly occupied and its contribution is integrated out to an effective field felt by the active space \cite{PhysRevX.10.011004} of remaining five orbitals, Fig.~\ref{fig:Intro}. The latter spans ten spin-orbitals, and as such the system can be described with a 10-qubit circuit. 

For RyRz and Ry ans\"atze, the effect of qubit permutations on circuit depth is shown in Figs.~\ref{fig:all_plots}a and ~\ref{fig:all_plots}e, respectively.  We note here that for both ans\"atze at depth $L = 2$, the VQE converged to a correlated wave function whose energy is comparable to the Hartree-Fock (HF) one. For depth $L = 4$, ca. 70\% of \textit{pseudo}-correlation energy $\Delta E_\mathrm{pc}$ (Fig.~\ref{fig:all_plots}a) is captured for the permuted Hamiltonian, where $\Delta E_\mathrm{pc}$ is defined as the difference between the Hartree-Fock (HF) energy $E_\mathrm{HF}$ and the exact $E_\mathrm{exact}$ ground state energy for reduced active space ($\Delta E_\mathrm{pc}$ = 12.2 kcal/mol). In contrast, for the unpermuted (default) qubit Hamiltonian, the VQE energy is still comparable to that of HF energy even for the depth $L = 8$, indicating that it generates insufficient entanglement of the qubits. Similar results are observed with Ry ansatz, Fig.~\ref{fig:all_plots}e.

\subsubsection{$\mathrm{H}_2$, $(\mathrm{H}_2)_2$ and $\mathrm{H}_4$ Molecular Species}

The improvement in energy convergence due to qubit permutations on 8-qubit circuit depth are well-observed for $\mathrm{H}_2$ (6-31G basis), non-covalently bound (Van der Waals) complex $(\mathrm{H}_2)_2$ (sto-3g basis), and $\mathrm{H}_4$ third-order saddle point (three imaginary frequencies in the Hessian matrix, sto-3g basis), Fig.~\ref{fig:all_plots}b-d and Fig.~\ref{fig:all_plots}f-h for RyRz and Ry ans\"atze, respectively. Note that the level of correlations in these molecular systems is expected to increase as follows: $\mathrm{H}_2< (\mathrm{H}_2)_2 \ll \mathrm{H}_4$. Consequently, smaller difference between permuted and unpermuted $E_\mathrm{VQE}$ can be observed in the corresponding panels. We also note here that $\mathrm{H}_4$ is so strongly correlated, that already at depth $L = 2$ converged VQE corresponds to a highly-correlated wave function even when unpermuted Hamiltonian is employed, $E_\mathrm{HF} - E_\mathrm{VQE}\mathrm{(unpermuted)}$ $\sim$ 100 kcal/mol, Fig.~\ref{fig:all_plots}d,h.

\subsubsection{$\mathrm{H}_3^+$ Cation}
Since VQE performs an unconstrained energy optimization in the Fock space of the original electronic problem, calculation of exact energy for charged molecules (ions) could be a challenging task \cite{ryabinkin2018constrained}. Indeed, as charge-neutral molecules are usually energetically more preferable over their ions, VQE will invariably collapse to a lower energy of the corresponding neutral form. Among other options \cite{ryabinkin2018constrained}, one could design quantum circuit which conserves the number of electrons, and force the search to be in the expected subspace \cite{Gard2020}.  

\begin{figure}[t]
    \centering
    \includegraphics[width=8.5cm]{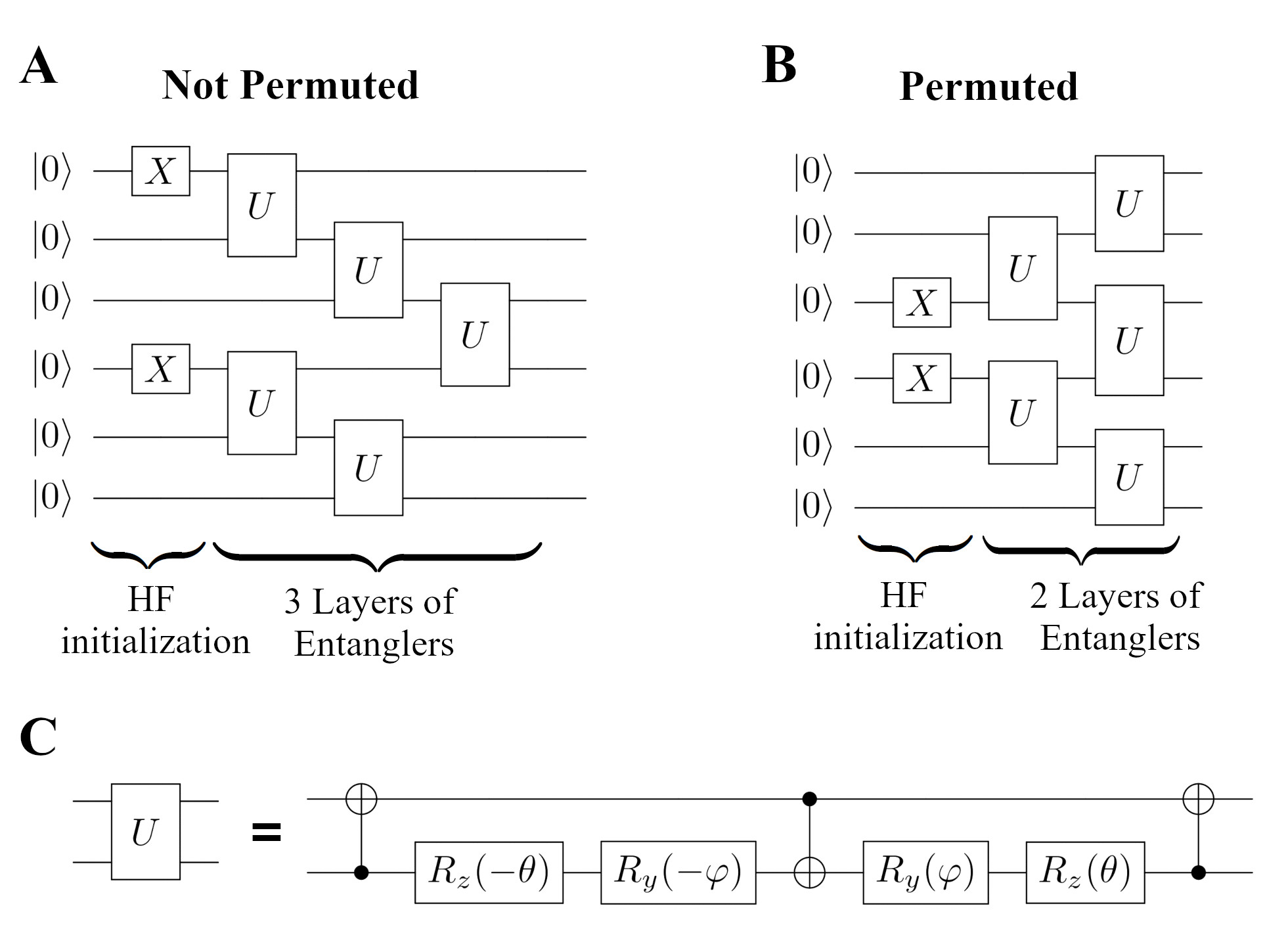}
    \caption{Example of particle preserving ans\"atze that were used in $H_3^+$ simulation. (A) 3-Depth ansatz used with unpermuted Hamiltonian; (B) 2-Depth ansatz used with permuted Hamiltonian; (C) Illustration of electron preserving entangler.}
    \label{fig:pp}
\end{figure}

\begin{figure}[t]
    \centering
    \includegraphics[width=7cm]{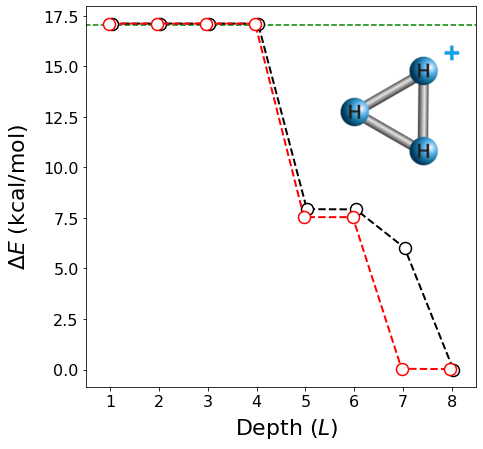}
    \caption{Energy error $\Delta E = E_\mathrm{VQE} - E_\mathrm{exact}$ versus ansatz depth $L$ for 6-qubit $\mathrm{H}_3^+$ cation based on default (black) and permuted qubit Hamiltonians (red) built from \textit{exact} wave function. Particle-preserving ans\"atze were used as shown in Fig.~\ref{fig:pp}. Green dotted-line corresponds to \textit{pseudo}-correlation energy defined as $\Delta E_\mathrm{pc} = E_\mathrm{HF} - E_\mathrm{exact}$.}
    \label{fig:h3_pp_graph}
\end{figure}

The JW mapping, which relates fixed-particle-number subspaces to the fixed-excitation subspaces in the qubit Hilbert space, seems to be the simplest to approach such a circuit. A complication, however, is a large number of two-qubit gates that act on states $\ket{00}$ or $\ket{11}$ in the HF reference state, in which case electron preserving gates must act as the identity. Thus, we remove these ineffective gates in our ans\"atze when reporting the depth of our circuit to get a more appropriate comparison between unpermuted vs permuted Hamiltonians. We display this result for our $\mathrm{H}_3^+$ cation in Fig.~\ref{fig:pp}a,b along with a decomposition of the electron-preserving gates, which we take from Ref. \cite{Gard2020}.  Note that the HF layer was initialized depending on the permutation used. Thus, if qubit 1 corresponds to an occupied orbital and after the permutation it swaps with unoccupied qubit 5, we initialize qubit 5 in a $\ket{1}$ state and qubit 1 in a $\ket{0}$ state. Using this approach, we again found that permutations help us to reduce the circuit depth required to reach lower $\Delta E$ although the effect is less profound compared to that for neutral molecules, Fig.~\ref{fig:h3_pp_graph}. At a higher depth $L = 8$, the results equalize.

\subsection{Molecular Systems with \textit{Approximate} Wave Functions (PermVQE)}

\begin{figure}[t]
    \centering
    \includegraphics[width=7cm]{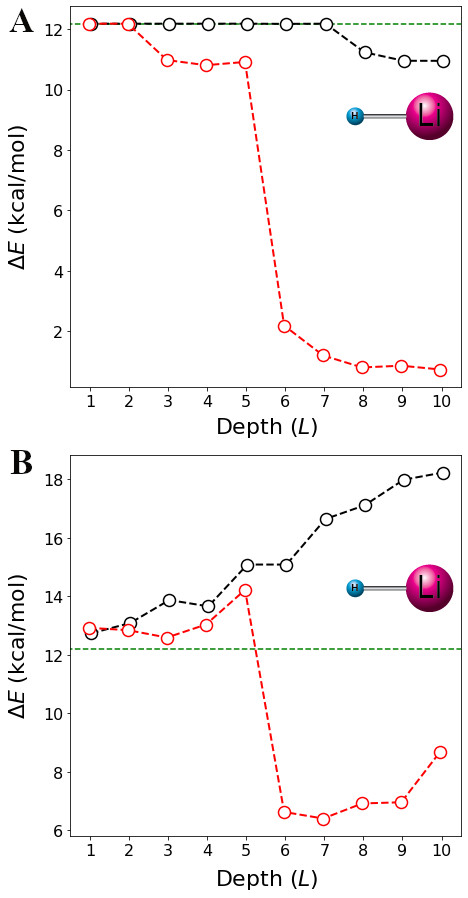}
    \caption{Energy error $\Delta E = E_\mathrm{VQE} - E_\mathrm{exact}$ versus Ry ansatz depth $L$ for 10-qubit LiH/sto-3g in a reduced active space based on default (black) and permuted qubit Hamiltonians (red) built from \textit{approximate} wave functions (up to three PermVQE iterations are used to build best entanglement maps). Results are based on \textbf{A}: Statevector (noiseless) simulator; \textbf{B} - Qasm (noisy) simulator. Green dotted-lines correspond to \textit{pseudo}-correlation energy defined as $\Delta E_\mathrm{pc} = E_\mathrm{HF} - E_\mathrm{exact}$.}
    \label{fig:lih_permvqe}
\end{figure}

So far we demonstrate the proof-of-concept beneficial effect of qubit permutations on circuit depth for toy Ising models and various molecular systems based on ideally permuted Hamiltonians where exact entanglement maps were built from the exact wave functions under noise-free conditions. We now describe PermVQE presented in Fig.~\ref{fig:algorithm} on the example of LiH molecule based on approximate wave function under noise-free and noisy conditions without resorting to exact solution. Each iteration uses the current best approximation to the ground state. Because von Neumann entropy $S$ is a function of density-matrix values, the approximate wave function for molecular systems must be at minimum correlated (\textit{e.g.} improved over uncorrelated HF wave function) before $S$ is calculated.

For LiH molecule in the reduced active space of sto-3g basis, some correlations are observed at $L = 1$ at $L = 2$ for Ry ansatz based on the entanglement maps shown in Fig. S6 in SM. Despite the apparent correction from $L = 1$ to $L = 2$, the system is not correlated enough for further improvement. As expected, upon increasing $L$ under noise-free conditions ($L \geq 3$), one can clearly see the profound effect of qubit permutations on the VQE converged energy, Fig. \ref{fig:lih_permvqe}A. For example, at $L = 7$, the converged VQE energy is at the level of chemical accuracy for the permuted Hamiltonian ($\Delta E$ = 1.2 kcal/mol). In contrast, the converged VQE energy is at the level of HF energy for the unpermuted case. We present the tabulated number of cost-function evaluations per PermVQE iteration for converged VQE per $L$ for unpermuted and permuted Hamiltonians, as well as selected PermVQE iterations for $L = 1, 2, 5$ and 10 (up to three for the system under studies) including corresponding JW entanglement maps in Table S2 and Fig. S7 in SM.

We also note the following empirical observation. One could intuitively expect that the approximate wave function to build the best-permuted Hamiltonian (within given $L$) should ideally have a reasonable overlap with the exact wave function. Results obtained for VQE based on unpermuted vs permuted Hamiltonian for $L = 2$ to 7 demonstrate a weak dependence on the accuracy of the initial wave function. The only relevant criterion seems to be that the variational state develops sufficient correlations, such that the correlation pattern is noticeable. Our experiments suggest that correlations do not have to be accurately computed for the purpose of finding favourable permutation of qubits. Indeed, the wave function is further refined in each iteration under fixed \textit{L}, e.g. see Fig. S7 in SM. 

Finally we study the effects of qubit permutations on noisy simulator. Optimized ansatz parameters under a noisy environment were used for the energy evaluation. No error mitigation was performed here and error rates were kept very low to obtain meaningful results, being $5 \times 10^{-5}$ for 1-qubit gates, $5 \times 10^{-4}$ for 2-qubit gates. The average energy values after 10 trials are shown in Fig. \ref{fig:lih_permvqe}B. For the unpermuted (default) Hamiltonian, the energy is not improved below the HF value and an increase of ansatz depth leads to a greater error in energy $\Delta E$ due to additional noise that those layers introduce. Similar behavior is observed for several initial points of permuted Hamiltonian ($L \leq 5$). However, a noticeable improvement is observed from $L \geq 6$ for the case of permuted Hamiltonian. After $L = 7$, the $\Delta E$ further increases as expected. We stress that the final permutation (for a given $L$) is obtained in several steps of the algorithm outlines in Fig.~\ref{fig:algorithm}. In each step, the current best approximation to the ground state  is used (as opposed to the exact ground state used for proof-of-concept calculations presented in previous sections).

\section{Conclusions}
In this work we show that encoding strongly interacting spin-orbitals of molecular systems into proximal qubits on a linear chain architecture quantum chip, naturally reduces the circuit depth needed to prepare the ground state for the quantum chemistry electronic structure problem. Assuming direct mapping between qubits and canonical spin-orbitals (eigenfunctions of the Fock operator), i.e. each qubit represents the occupation number of a particular spin-orbital, the two-qubit mutual information describes the entanglement between two individual qubits, whereas JW encoding is the most-suitable for permuting qubits. With this observation, we developed a PermVQE algorithm which uses default (unpermuted) Hamiltonians generated with IBM's Qiskit or Google's Openfermion frameworks. PermVQE then iteratively converts inputs into permuted Hamiltonians based on mutual information from approximate wave function, and performs VQE with improved precision at a given circuit depth $L$. We remark that a different approach based on mutual information maximization to select entangling ans\"atze (rather than our mutual information minimization to select permutations) was reported in Ref.~\cite{zhang2020mutual} during the preparation of this manuscript.

We believe that the proposed method will facilitate the simulation of larger molecular systems with NISQ devices by reducing the depth length, and contribute to the demonstration of chemical advantage. Furthermore, we remark that our correlation-informed permutation approach can be combined within the other variational quantum algorithms beyond VQE, for example, to reduce ansatz depth in variational quantum algorithms for simulating dynamics~\cite{cirstoiu2019variational,commeau2020variational,yuan2019theory,endo2018variational}, solving linear systems~\cite{bravo-prieto2019,Xiaosi}, and compiling~\cite{QAQC,sharma2019noise}.

Finally, we note that PermVQE can be employed for any hardware connectivity. Future study might include extending the permutation approach to take advantage of more highly connected qubit architectures, such as grid or hexagonal lattices, which would require modified and more complicated cost functions.

\appendix

\section{Appendix: Ansatz Methods}

Hardware-efficient RyRz ansatz that was built on the basis of 
IBM's Qiskit software package \citep{Abraham2019}, was used to prepare quantum states for Ising toy models, whereas partially modified RyRz and Ry ans\"atze were used to prepare quantum states for neutral molecules. For RyRz, additional Ry and Rz rotational final layers were included to provide more flexibility to the parameterized circuit,  Fig. S2 in SM. For Ry, all Rz gates were removed and additional final Ry rotational layer was included, see Fig. S3 in SM. This ansatz can be used since all ground states of systems that we analyze possess time-reversal symmetry. Thus, we expect all state coefficients to be purely real, eliminating the need for Z rotations. Electron-preserving ansatz shown in Fig.~\ref{fig:pp} was used to model $\mathrm{H}_3^+$ cation.

\section{Appendix: Computational Methods}

All quantum simulations were performed using the Qiskit 0.19.6 package \citep{Abraham2019}. For the VQE and noisy VQE, the Qiskit's Statevector and Qasm simulators were used, respectively. For a classical optimization, the COBYLA \citep{Powell1998} protocol was used with 200000 maximum iterations for investigated molecular systems. The lowest eigenvalue was obtained from multiple trial runs (5-10) for each depth of the ansatz. For model Ising Hamiltonians, the same classical optimizer was used with maximum 10000 iterations. The PermVQE calculations with approximate reference wave functions were performed with the same classical optimizer with maximum 100000 iterations and maximum allowed 3 consecutive permutations of the corresponding Hamiltonians. The noise model was incorporated by running the final step of VQE calculation using qasm simulator with 10000 shots per Pauli-word evaluation and the following error rates: $5 \times 10^{-5}$ for 1-qubit gates, $5 \times 10^{-4}$ for 2-qubit gates. The geometries of molecular systems were initially preoptimized (in the gas phase) at the Hartree-Fock or Full Configurational Interaction level for $(\mathrm{H}_2)_2$ using electronic structure Gaussian~16 (Revision B.01) package \citep{Frisch2016}. Table S3 in SM contains additional information (xyz coordinates, energy data). The molecular orbitals were visualized using ChemCraft software (https://www.chemcraftprog.com/). The entanglement maps for each run were built using the exact or approximate wave function of a given system. The brute force approach was applied to find the best permutation of a given entanglement map.

\section*{Acknowledgements}

Research presented in this article was supported by the Laboratory Directed Research and Development (LDRD) program of Los Alamos National Laboratory (LANL) under project number 20200056DR. LANL is operated by Triad National Security, LLC, for the National Nuclear Security Administration of U.S. Department of Energy (contract no. 89233218CNA000001). We thank LANL Institutional Computing (IC) program for access to HPC resources. JS acknowledges support from the U.S. Department of Energy (DOE) through a quantum computing program sponsored by the LANL Information Science \& Technology Institute. This work was conducted in part at the Center for Integrated Nanotechnologies, a U.S. Department of Energy, Office of Basic Energy Sciences user facility. PJC and LC were partially supported by the U.S. Department of Energy (DOE), Office of Science, Office
of Advanced Scientific Computing Research, under the
Accelerated Research in Quantum Computing (ARQC)
program.

\newcounter{ct}
\forloop{ct}{1}{\value{ct} < 11}%
{%
  \clearpage
  

\section*{Supplementary material}

# Supplemental Material

# Correlation-Informed Permutation of Qubits for Reducing Ansatz Depth in VQE

Nikolay V. Tkachenko,[1] James Sud,[2] Yu Zhang,[3,*] Sergei Tretiak,[3] Petr M. Anisimov,[4] Andrew T. Arrasmith,[3] Patrick J. Coles,[3] Lukasz Cincio,[3,†] and Pavel A. Dub[1,‡]

[1] Chemistry Division, Los Alamos National Laboratory, Los Alamos, New Mexico 87545, USA

[2] Department of Physics, University of California, Berkeley, CA 94720, USA

[3] Theoretical Division, Los Alamos National Laboratory, Los Alamos, New Mexico 87545, USA

[4] Accelerators and Electrodynamics Group, Los Alamos National Laboratory, Los Alamos, New Mexico 87545, USA

[*] Corresponding author: zhy@lanl.gov.

[†] Corresponding author: lcincio@lanl.gov.

[‡] Corresponding author: pdub@lanl.gov.

# Supplementary Figures

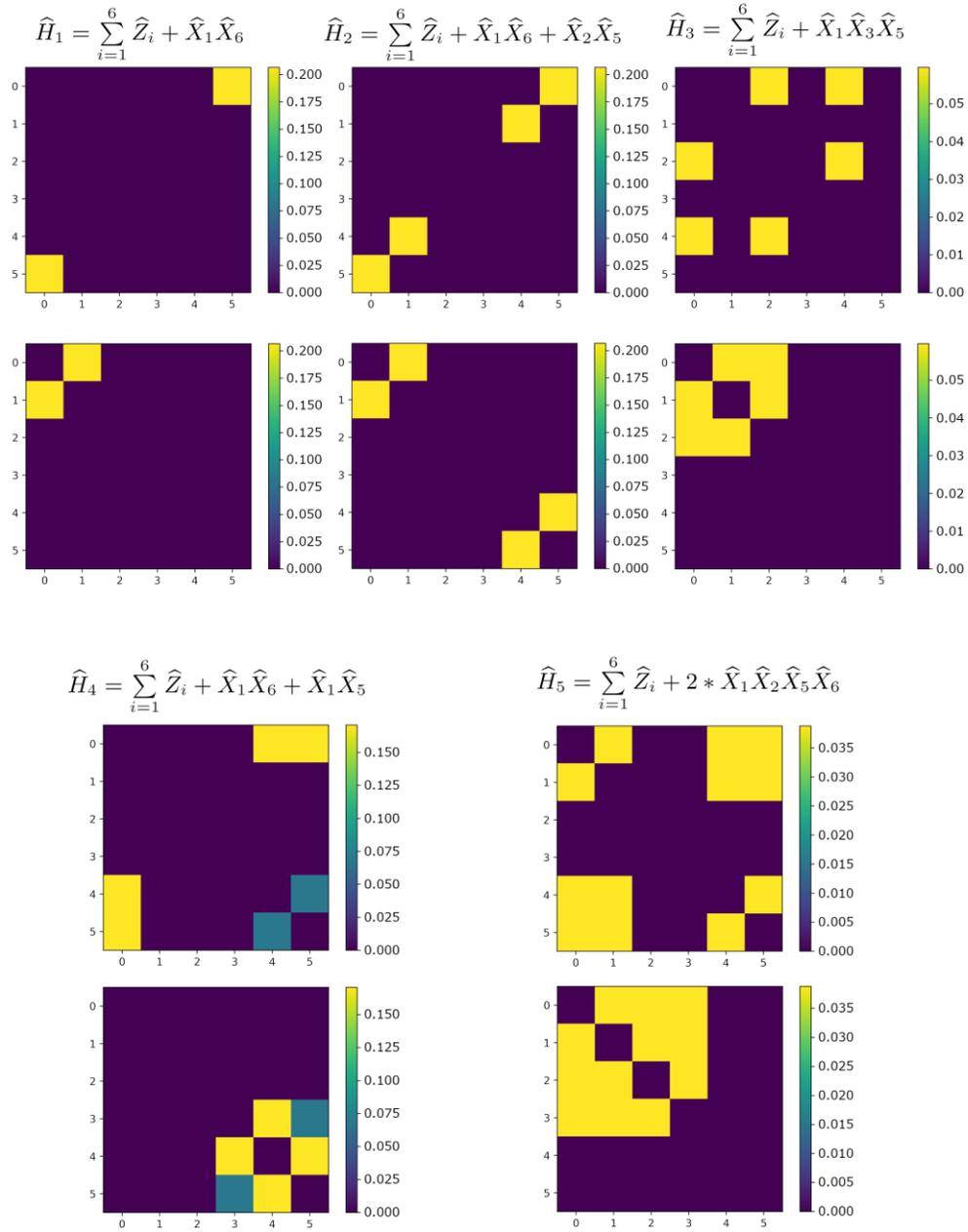

**Figure S1.** Entanglement maps built from exact wave function for investigated toy Ising Hamiltonians. Top plot: before permutation, Bottom plot: after permutation. Numbers on the axes indicate qubit indices. Different values of mutual information $I_{ij}$ are shown with different colors.

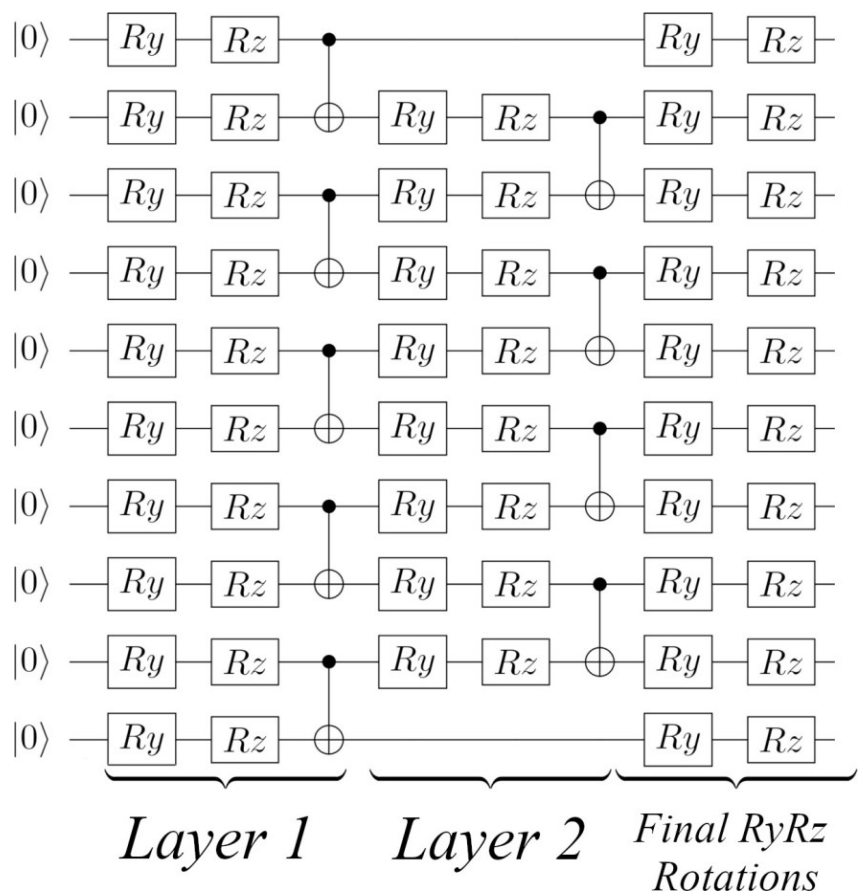

**Figure S2.** llustration of the 10-qubit 2-Depth RyRz Ansatz that was used for VQE calculations of molecular systems.

**Figure S3.** llustration of the 10-qubit 4-Depth Ry Ansatz that was used for VQE calculations of molecular systems.

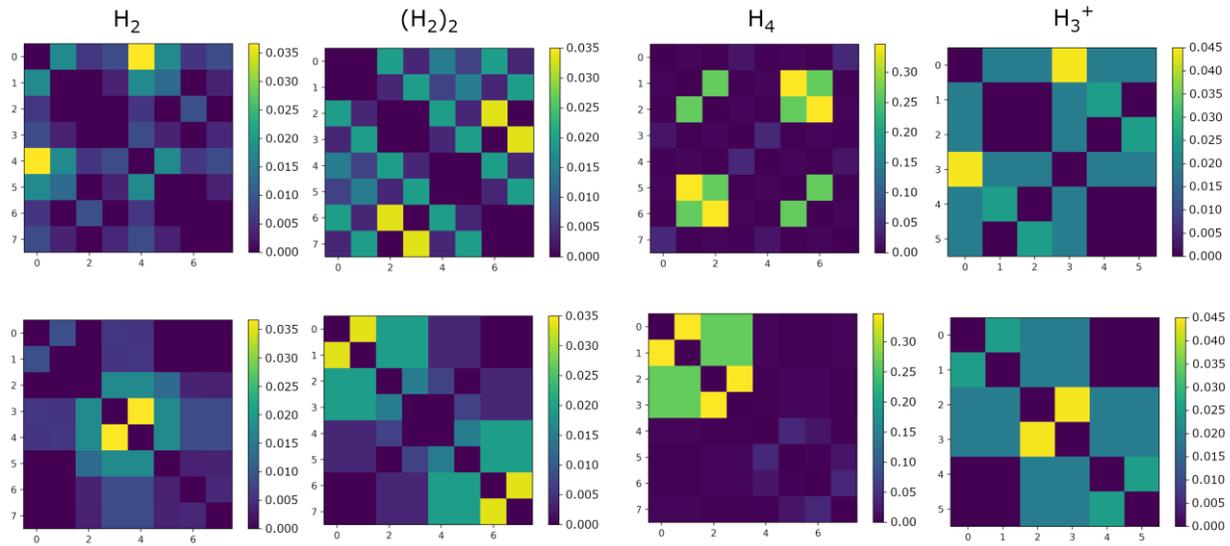

**Figure S4.** Entanglement maps built from exact wave function for investigated molecular systems in Jordan-Wigner mapping. Top row: before permutation, Bottom row: after permutation. Numbers on the axes indicate qubit indices. Different values of mutual information $I_{ij}$ are shown with different colors.

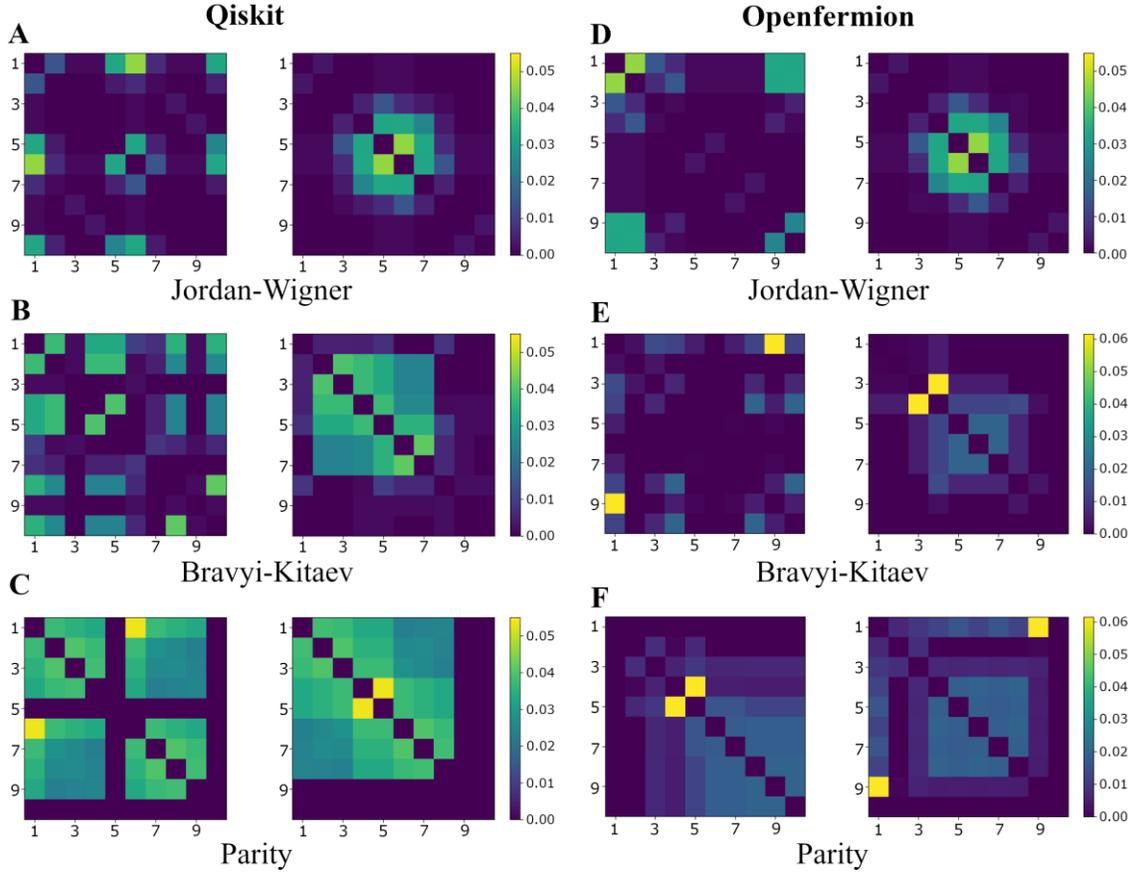

**Figure S5.** Comparison of entanglement maps built from exact wave function for LiH/sto-3g in a reduced active space of 10 spin-orbitals in three different mappings obtained from Qiskit (A-C) and Openfermion (D-E) packages. Left column: before permutation, Right column: after permutation. Numbers on the axes indicate qubit indices. Different values of mutual information $I_{ij}$ are shown with different colors.

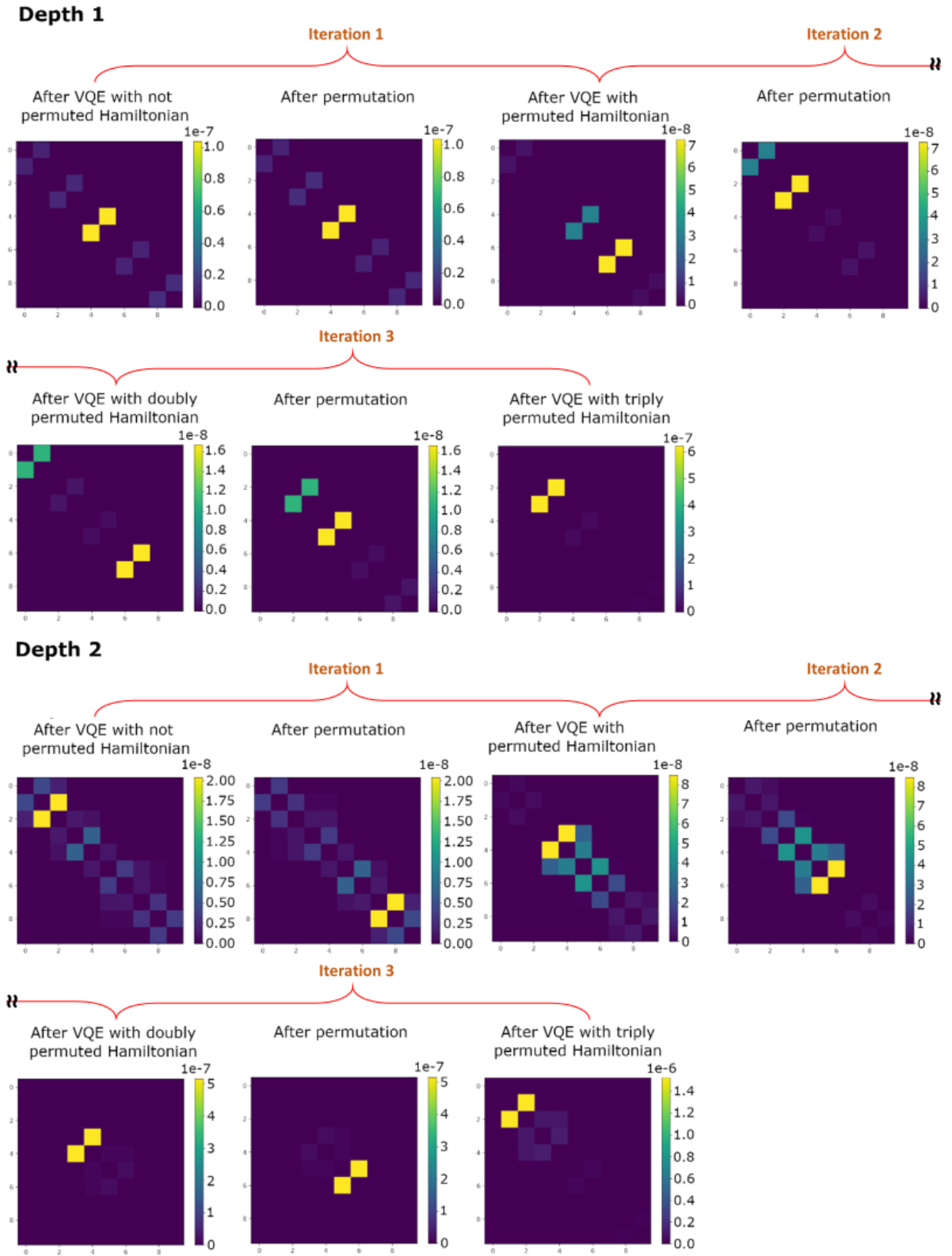

**Figure S6.** Entanglement maps built during PermVQE procedure from a wave function obtained after VQE calculations using Ry ansatz with Depth 1 and 2. Numbers on the axes indicate qubit indices. Different values of mutual information $I_{ij}$ are shown with different colors.

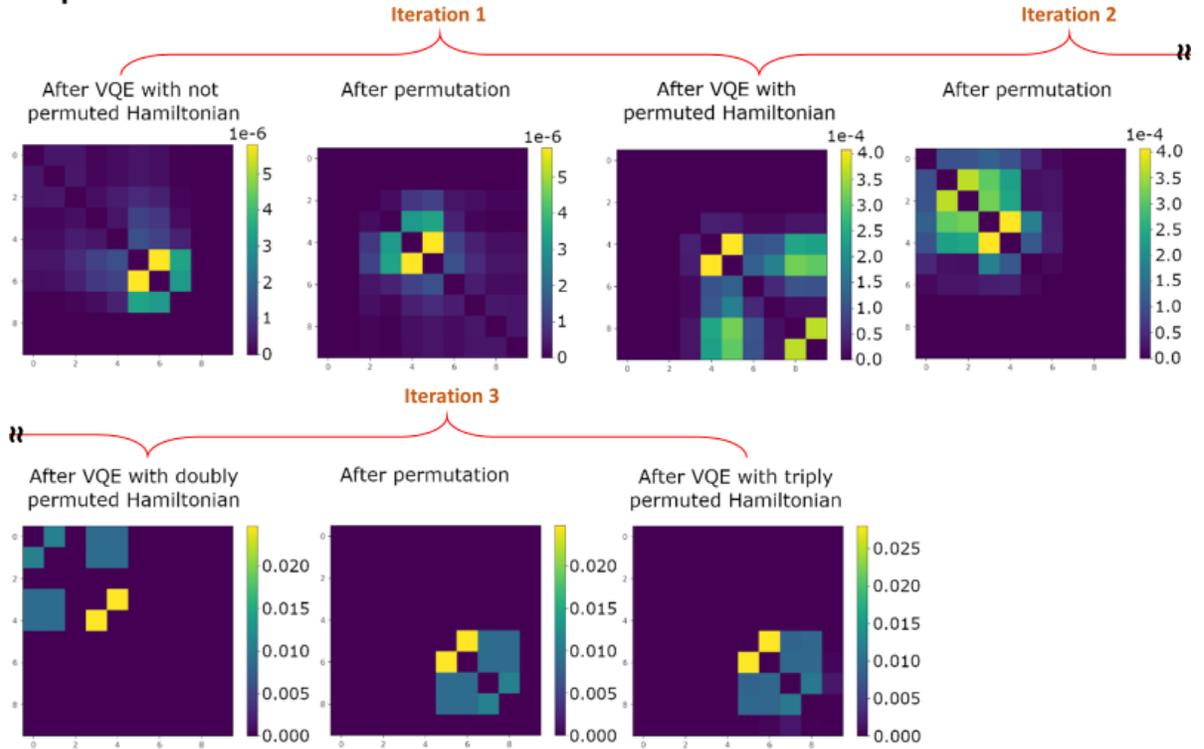
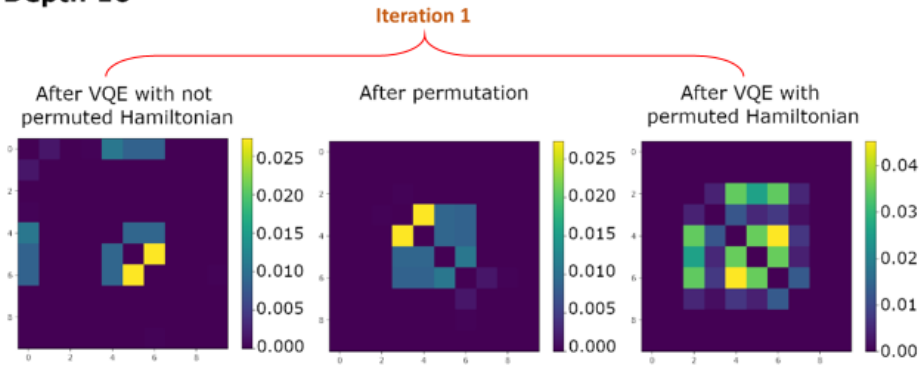

**Figure S7.** Entanglement maps built during PermVQE procedure from a wave function obtained after VQE calculations using Ry ansatz with Depth 5 and 10. Numbers on the axes indicate qubit indices. Different values of mutual information $I_{ij}$ are shown with different colors.

# Supplementary Tables

**Table S1.** The comparison of the difference in cost-function values with the difference in depth of the corresponding ansatzes for each model Hamiltonian.

| Hamiltonian | Difference in ansatz depth | Difference in cost function |
|---|---|---|
| $\hat{H}_1$ | 6 | 19.84 |
| $\hat{H}_2$ | 7 | 26.45 |
| $\hat{H}_3$ | 2 | 4.30 |
| $\hat{H}_4$ | 5 | 25.81 |
| $\hat{H}_5$ | 0 | 19.84 |

**Table S2.** Number of cost-function evaluations during the PermVQE calculations of LiH molecule.

| Depth | Not Permuted | Permuted |
|---|---|---|
| 1 | 1000-3000 | 800-6100 |
| 2 | 1100-9000 | 1200-6500 |
| 3 | 1200-5400 | 37800 |
| 4 | 7700-100000 | 100000 |
| 5 | 13000-41400 | 100000 |
| 6 | 31000-100000 | 100000 |
| 7 | 48000-72290 | 100000 |
| 8 | 100000 | 100000 |
| 9 | 100000 | 100000 |
| 10 | 100000 | 100000 |

**Table S3.** Cartesian coordinates and corresponding energies of molecular systems that were used in current study.

| | |
|---|---|
| LiH | Basis set: STO-3G<br>Hartree-Fock Energy: -7.86311647 Hartree<br>FCI Energy: -7.88253781 Hartree |
| | 3    0.000000000    0.000000000    0.000000000<br>1    0.000000000    0.000000000    1.547220000 |
| $H_2$ | Basis set: 6-31G<br>Hartree-Fock Energy: -1.12682783 Hartree<br>FCI Energy: -1.15152019 Hartree |
| | 1    0.000000000    0.000000000   -0.364980000<br>1    0.000000000    0.000000000    0.364980000 |
| $H_4$ | Basis set: STO-3G<br>Hartree-Fock Energy: -1.71135540 Hartree<br>FCI Energy: -1.95620001 Hartree |
| | 1    0.000000000    0.000000000    0.792141000<br>1    0.792141000    0.000000000    0.000000000<br>1    0.000000000    0.000000000   -0.792141000<br>1   -0.792141000    0.000000000    0.000000000 |

| (H$_2$)$_2$ | Basis set: STO-3G |   |   |   |
|---|---|---|---|---|
|   | Hartree-Fock Energy: -2.23401601 Hartree |   |   |   |
|   | FCI Energy: -2.27461770 Hartree |   |   |   |
|   | 1 | 0.000000000 | 0.367436000 | -2.126445000 |
|   | 1 | 0.000000000 | -0.367436000 | -2.126445000 |
|   | 1 | 0.000000000 | 0.000000000 | 1.759009000 |
|   | 1 | 0.000000000 | 0.000000000 | 2.493881000 |
| H$_3^+$ | Basis set: STO-3G |   |   |   |
|   | Hartree-Fock Energy: -1.24686001 Hartree |   |   |   |
|   | FCI Energy: -1.27414447 Hartree |   |   |   |
|   | 1 | 0.000000000 | 0.558243000 | 0.000000000 |
|   | 1 | 0.483452000 | -0.279121000 | 0.000000000 |
|   | 1 | -0.483452000 | -0.279121000 | 0.000000000 |


}

\end{document}